# Stability and Robustness of Disturbance Observer Based Motion Control Systems

Emre Sariyildiz, *Student Member IEEE*, Kouhei Ohnishi, *Fellow, IEEE*

*Abstract*—This paper analyzes the robustness and stability of a disturbance observer (DOB) and a reaction torque observer (RTOB) based robust motion control systems. Conventionally, a DOB is analyzed by using an ideal velocity measurement that is obtained without using a low-pass-filter (LPF); however, it is impractical due to noise constraints. A LPF of velocity measurement changes the robustness of a DOB significantly and puts a new design constraint on the bandwidth of a DOB. A RTOB, which is used to estimate environmental impedance, is an application of a DOB. The stability of a RTOB based robust force control system has not been reported yet, since its oversimplified model is derived by assuming that a RTOB has a feed-forward control structure. However, in reality, it has a feed-back control structure; therefore, not only the performance, but also the stability is affected by the design parameters of a RTOB. A new practical stability analysis method is proposed for a RTOB based robust force control system. Besides that novel and practical design methods, which improve the robustness of a DOB and the stability and performance of a RTOB based robust force control system, are proposed by using the new analysis methods. The validity of the proposals is verified by simulation and experimental results.

## I. Introduction

IN the motion control field, two-degrees-of-freedom (2-DOF) controllers, in which the robustness and performance of servo-systems are controlled independently, have long been used due to their significant advantages, e.g., the robustness can be achieved by using low gain controllers [1- 4]. Several design methods have been proposed to achieve a 2-DOF controller such as generalized internal model control (GIMC) and DOB based control methods [5, 6, and 7]. Among them, a DOB is one of the most popular robust control tools, since the robustness can be adjusted in a desired bandwidth, intuitively [7].

A DOB, which was proposed by Ohnishi et al., has been used in several motion control applications, e.g., robotics, industrial automation, automotive, etc., due to its simplicity and efficiency [7-12]. It estimates external disturbances and system uncertainties, and the robustness of a system is achieved by feeding back the estimated disturbances in an inner-loop [6, 13]. To achieve performance goals, an outer-loop controller is designed by considering only the nominal plant model, e.g., acceleration based control [7, 14, and 15]. The bandwidth of a DOB and the nominal inertia and torque coefficient of a motor are the fundamental design parameters of a DOB based robust motion control system. The bandwidth of a DOB is desired to set as high as possible to estimate/suppress disturbances in a wide frequency range; however, it is limited by noise and robustness constraints [16, 17]. Several researches have been conducted to improve the bandwidth of a DOB by suppressing the noise of velocity measurement [18, 19, and 20]. To suppress the noise, velocity is generally estimated by using a LPF in the design of a DOB; however, so far, it has not been considered to simplify the analyses. It is a well-known fact that the stability and performance of a DOB based robust motion control system can be improved by using higher/lower nominal inertia/torque coefficient; however, its upper/lower bound has not been shown yet [21]. In reality, the nominal plant parameters are limited by the robustness of a DOB and practical design constraints, so the stability and performance cannot be improved freely.

A RTOB, which was proposed by Murakami et al., is an application of a DOB and is used to estimate environmental impedance [22]. It is simply designed by subtracting system uncertainties from the input of a DOB [22]. Although a DOB and a RTOB are quite similar, only the latter has a model based control structure that is the main challenging issue in its design. Several superiorities of a RTOB over a force sensor, e.g., higher force control bandwidth, stability improvement, force-sensorless force control, etc., have been shown experimentally in the literature [22, 23, 24, and 25]. In the conventional analysis and design methods of a RTOB based robust force control system, oversimplified methods, which consider only the performance, are used for the sake of simplicity, and it is designed by using the DOB's design parameters, e.g., the bandwidths of a DOB and a RTOB are set to the same value in general [22, 23]. However, in reality, not only the performance, but also the stability changes significantly by the design parameters of a RTOB, since it has a feed-back control structure. Therefore, the stability of a RTOB based robust force control system should be analyzed.

Manuscript received September 20, 2013; revised December 7, 2013 and January 19, 2014; accepted April 19, 2014.

This research was supported in part by the Ministry of Education, Culture, Sports, Science, and Technology of Japan under Grant- in-Aid for Scientific Research (S), 25220903, 2013.

E. Sariyildiz and K. Ohnishi are with the Ohnishi Laboratory, Department of System Design Engineering, Keio University, Yokohama, 223-8522, Japan. (e-mail:emre@sum.sd.keio.ac.jp, ohnishi@sd.keio.ac.jp)

In this paper, new robustness and stability analysis methods are proposed for the DOB based robust motion control systems. The LPF of velocity measurement is considered, and it is shown that there is a trade-off between the robustness and stability in the design of a DOB; the bandwidth of a DOB and nominal inertia/torque coefficient have upper/lower bounds due to the robustness constraint. A new design criterion is proposed to adjust the trade-off between the robustness and stability. Besides that a new stability analysis is proposed for a RTOB based robust force control system. It is shown that the stability of the robust force control system is affected significantly by the design parameters of a DOB and a RTOB. A DOB and a RTOB can be designed as a phase lead-lag compensator by setting their bandwidths to different values, and the stability and performance of the robust force control system can be improved by increasing the bandwidth of a RTOB. The identified inertia in the design of a RTOB changes the stability of the robust force control system: if the identified inertia is higher than exact inertia, then there is a right half plane zero in the open loop transfer function of the robust force control system so the stability and performance deteriorate significantly. New design criteria are proposed to improve the stability and performance of a RTOB based robust force control system. Authors have recently proposed a new adaptive design method for the RTOB based robust force control systems by using the proposed analysis methods [26].

The rest of the paper is organized as follows. In section II, a DOB and a RTOB are presented briefly. In section III, the stabilities of DOB based robust position and force control systems are analyzed, and new design criteria are proposed. In section IV, simulation and experimental results are given. The paper ends with conclusion given in the last section.

## II. DISTURBANCE AND REACTION TORQUE OBSERVERS

### A. Disturbance Observer

A DOB, which is shown in Fig. 1, estimates external disturbances and system uncertainties such as external load, friction, inertia variation, and so on. The robustness of the system is achieved by feeding-back the estimated disturbances as shown in the figure. If it is assumed that $g_v$ is infinite, i.e., ideal velocity measurement is achieved, then the dynamic equations of the robust motion control system are derived from Fig. 1 as follows:

$$K_{\tau n} I_m - J_{mn} \ddot{q}_m = \tau_m^{dis} = \tau_m^d + \Delta J_m \ddot{q}_m - \Delta K_\tau I_m \quad (1)$$

$$\hat{\tau}_m^{dis} = \frac{g_{DOB}}{s + g_{DOB}} \tau_m^{dis} \quad (2)$$

where

$J_m, J_{mn}$    Uncertain and nominal inertias;
$K_\tau, K_{\tau n}$    Uncertain and nominal torque coefficients;
$I_m, I_m^{des}, I_m^{cmp}$    Total, desired and compensate motor currents;
$q_m, \dot{q}_m, \ddot{q}_m$    Angle, velocity and acceleration;
$\ddot{q}_m^{des}$    Desired acceleration;
$\dot{q}_m^{noise}$    Noise of velocity measurement;
$g_{DOB}$    Cut-off frequency of DOB;
$g_v$    Cut-off frequency of velocity measurement;
$\Delta J_m = (J_m - J_{mn})$    Inertia variation;
$\Delta K_\tau = (K_\tau - K_{\tau n})$    Torque coefficient variation;
$\tau_m^{load}$    Loading torque;
$\tau_m^{frc}$    Friction torque;
$\tau_m^{int}$    Interactive torque;
$\tau_m^d = \tau_m^{int} + \tau_m^{load} + \tau_m^{frc}$    Total external disturbance;
$\tau_m^{dis}, \hat{\tau}_m^{dis}$    Total system disturbance and its estimation;

Equation (2) shows that a disturbance can be estimated precisely if it stays within the bandwidth of a DOB.

A DOB based robust motion control system has a multi-input single-output (MISO) control structure, and its transfer function is described as follows:

- If $g_v$ is infinite, then

$$\ddot{q}_m = \alpha \frac{s + g_{DOB}}{s + \alpha g_{DOB}} \ddot{q}_m^{des} - \frac{1}{J_m} T_{SEN}^{DOB} \tau_m^d - T_{CoSEN}^{DOB} s \dot{q}_m^{noise} \quad (3)$$

where $T_{SEN}^{DOB} = \frac{1}{1 + L_{DOB}(s)}$ and $T_{CoSEN}^{DOB} = \frac{L_{DOB}(s)}{1 + L_{DOB}(s)}$ are the sensitivity and co-sensitivity transfer functions, in which $L_{DOB}(s) = \alpha \frac{g_{DOB}}{s}$; and $\alpha = \frac{J_{mn} K_\tau}{J_m K_{\tau n}}$.

- If $g_v$ is finite, then

$$\ddot{q}_m = \alpha \frac{(s + g_v)(s + g_{DOB})}{s^2 + g_v s + \alpha g_v g_{DOB}} \ddot{q}_m^{des} - \frac{1}{J_m} T_{SEN}^{DOB} \tau_m^d - T_{CoSEN}^{DOB} s \dot{q}_m^{noise} \quad (4)$$

where $T_{SEN}^{DOB}$ and $T_{CoSEN}^{DOB}$ are same as defined above; however,

$$L_{DOB}(s) = \alpha \frac{g_v g_{DOB}}{s(s + g_v)}.$$

Equations (3) and (4) show that a DOB works as a phase lead-lag compensator that is adjusted by $\alpha$. If $\alpha > 1$, then a DOB works as a phase lead compensator, and the stability and performance are improved by increasing $\alpha$.

It is a well-known fact that a DOB requires precise velocity measurement [27]. In practice, $g_v$ should be finite to suppress noise and obtain precise velocity measurement in a determined bandwidth. Although it has never been considered so far, the robustness of a DOB changes significantly when a LPF is used in velocity measurement. It can be explained briefly as follows:

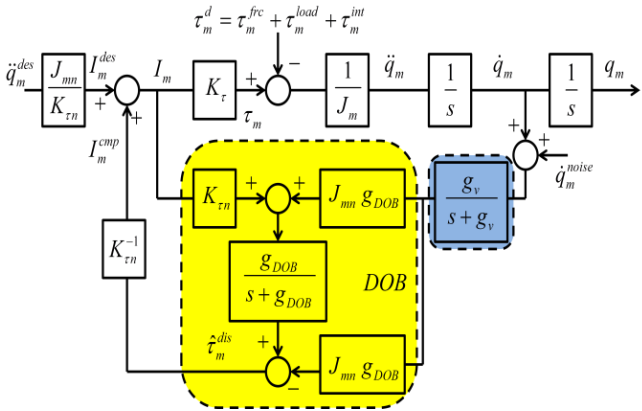

Fig. 1 A block diagram of a DOB

The sensitivity and co-sensitivity transfer functions, i.e., disturbance and noise responses, of a DOB based motion control system are given in (3) and (4). The relative degree of $L_{DOB}(s)$ is one and two when $g_v$ is infinite and finite, respectively. The Bode integral theorem shows that if the relative degree of $L_{DOB}(s)$ is higher than one, then $T_{SEN}^{DOB}$ cannot be shaped freely; the peak of $T_{SEN}^{DOB}$ at high frequencies increases as the sensitivity reduction at low frequencies is increased [17, 28]. Consequently, as shown in (4), $\alpha$ and $g_{DOB}$ cannot be increased freely due to the robustness constraint. A simple derivation for the robustness constraints of $\alpha$ and $g_{DOB}$ can be obtained as follows:

Let us consider the characteristic polynomial of $T_{SEN}^{DOB}/T_{CoSEN}^{DOB}$ and apply $g_v = \kappa g_{DOB}$. Then,

$$C_h(s) = s^2 + \kappa g_{DOB} s + \alpha \kappa g_{DOB}^2 \quad (5)$$

The natural frequency and damping coefficient of (5) are as follows:

$$w_n = \sqrt{\alpha\kappa}\, g_{DOB} \text{ and } \xi = 0.5\sqrt{\kappa\alpha^{-1}} \quad (6)$$

To improve the robustness of a DOB, i.e., to suppress the peak of $T_{SEN}^{DOB}/T_{CoSEN}^{DOB}$, if it is assumed that $\xi \geq 0.707$, then [29]

$$\alpha g_{DOB} \leq \frac{g_v}{2} \quad (7)$$

Equation (7) is a new design constraint which provides good robustness when a LPF is used in velocity measurement. The robustness of a DOB can be improved by increasing the lower constraint of $\xi$; however, the upper bounds of $\alpha$ and/or $g_{DOB}$ become more severe, so the stability and performance of a DOB deteriorate. Therefore, there is a trade-off between the robustness and stability and performance of a DOB based motion control system.

### B. Reaction Torque Observer

A RTOB, which is shown in Fig. 2, is used to estimate environmental impedance [22]. In this figure, $\hat{\tau}_m^{frc}$ and $\hat{\tau}_m^{int}$ denote estimated friction and interactive torques, respectively; $\Delta\hat{J}_m$ and $\Delta\hat{K}_\tau$ denote estimated inertia and torque coefficient variations, respectively; and $g_{RTOB}$ denotes the cut-off frequency of RTOB. Fig. 1 and 2 show that a DOB and a RTOB have quite similar control structures; however, only the latter is a model based control method.

### III. ROBUST POSITION AND FORCE CONTROL SYSTEMS

In this section, stability analyses of the robust position and force control systems will be presented.

#### A. Position Control

A block diagram of an acceleration based robust position control system is shown in Fig.3 [7]. In this figure, $q_m^{ref}$ and $\ddot{q}_m^{ref}$ denote angle and acceleration reference inputs, respectively; and $K_P$ and $K_D$ denote the proportional and derivative gains of the outer-loop controller, respectively. The other parameters are same as defined above. The transfer functions between $\ddot{q}_m^{ref}$ and $\ddot{q}_m$ are derived from Fig. 3 as follows:

$$\frac{\ddot{q}_m}{\ddot{q}_m^{ref}} = \frac{\alpha(s+g_{DOB})(s^2+K_D s+K_P)}{s^2(s+\alpha g_{DOB}) + \alpha(s+\alpha g_{DOB})(K_D s+K_P)} \quad (8)$$

when $g_v$ is infinite; and

$$\frac{\ddot{q}_m}{\ddot{q}_m^{ref}} = \frac{\alpha(s+g_v)(s+g_{DOB})(s^2+K_D s+K_P)}{s^2(s^2+g_v s+\alpha g_v g_{DOB}) + \alpha(s+g_v)(s+g_{DOB})(K_D s+K_P)} \quad (9)$$

when $g_v$ is finite. The characteristic functions of the robust position control system depend on $g_{DOB}, \alpha, K_P, K_D$ and $g_v$.

For the sake of simplicity, let us consider (8). If the stability of the transfer function is analyzed by using the Routh-Hurwitz theorem, then

$$\alpha^{-1} < 1 + g_{DOB}\frac{K_D}{K_P} + \frac{K_D}{g_{DOB}} + \frac{K_D^2}{K_P} \quad (10)$$

is derived as a stability criterion [29, 30]. It clearly shows that the stability is improved by using higher/lower nominal inertia/torque coefficient, i.e., increasing $\alpha$. However, as it is shown in (7), $\alpha$ cannot be increased freely due to the robustness constraint of a DOB. Therefore, there is a trade-off between the robustness of a DOB and the stability of a DOB based robust position control system.

Although, in general, it is assumed that the robustness and performance controllers are independent from each other, it is not true. The robustness of the position control system depends not only on the DOB, but also on the outer-loop performance controller. It can be shown as follows:

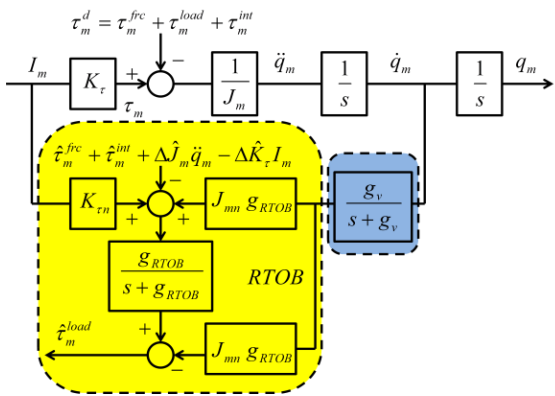

Fig. 2 A block diagram of a RTOB

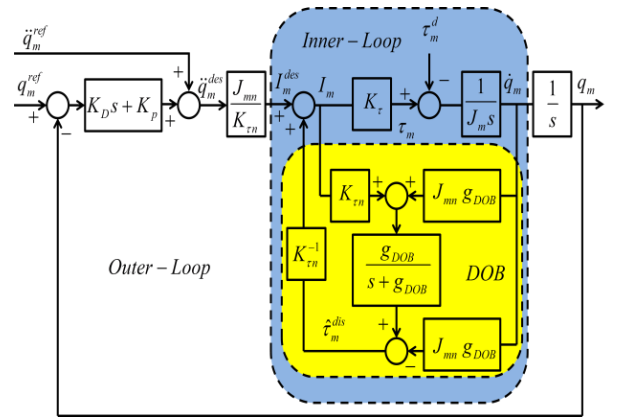

Fig. 3 A Block diagram of a DOB based robust position control system

The sensitivity and co-sensitivity transfer functions of the robust position control system are derived as follows:

$$T_{SEN}^{PC} = \frac{1}{1+L_{PC}(s)} \quad \text{and} \quad T_{CoSEN}^{PC} = \frac{L_{PC}(s)}{1+L_{PC}(s)} \quad (11)$$

where

$$L_{PC}(s) = \alpha \frac{g_{DOB}s^2 + (s+g_{DOB})(K_D s + K_P)}{s^3} \quad (12)$$

when $g_v$ is infinite.

$$L_{PC}(s) = \alpha \frac{g_v g_{DOB}s^2 + (s+g_v)(s+g_{DOB})(K_D s + K_P)}{s^3(s+g_v)} \quad (13)$$

when $g_v$ is finite.

Equation (11) shows that the robustness of the position control system can be improved by increasing the outer-loop control gain when $\alpha g_{DOB} > 0.5 g_v$; however, a DOB is still sensitive to noise and disturbances at high frequencies in the inner-loop.

### B. Force Control

A block diagram of a RTOB based robust force control system is shown in Fig. 4 [22]. In this figure, $C_f$ denotes the outer-loop force control gain, and the other parameters are same as defined above. A DOB provides the robustness of the force control system by estimating external disturbances and system uncertainties in the inner-loop. However, system uncertainties should be identified precisely to design a RTOB in the outer-loop. In the design of a RTOB, imperfect identification of system uncertainties may degrade not only the performance, but also the stability. The stability of the robust force control system is analyzed as follows:

Let us define the external load, i.e., environmental contact, by using a lumped spring-damper model as follows:

$$\tau_m^{load} = D_{env}(\dot{q}_m - \dot{q}_{env}) + K_{env}(q_m - q_{env}) \quad (14)$$

where $D_{env}$ and $K_{env}$ denote the environmental damping and stiffness coefficients, respectively; and $q_{env}$ and $\dot{q}_{env}$ denote the position and velocity of environment at equilibrium, respectively [31].

The dynamic equation of a RTOB is derived directly from Fig. 2 as follows:

$$\left(\hat{K}_\tau I_m - \hat{J}_m \ddot{q}_m - \hat{\tau}_m^{frc} - \hat{\tau}_m^{int}\right)\frac{g_{RTOB}}{s+g_{RTOB}} = \hat{\tau}_m^{load} \quad (15)$$

The transfer function between $\tau_{ref}^{load}$ and $\hat{\tau}_m^{load}$ is derived by using (1), (2), (15) and Fig. 4 as follows:

$$\frac{\hat{\tau}_m^{load}}{\tau_{ref}^{load}} = \frac{L_{RTOB}(s)}{1+L_{RTOB}(s)} \quad (16)$$

where $L_{RTOB}(s) = C_f \dfrac{g_{RTOB}\dfrac{J_{mn}}{K_{\tau n}}(s+g_{DOB})\varphi(s)}{s\{J_m s(s+\alpha g_{DOB})+(D_{env}s+K_{env})\}(s+g_{RTOB})}$ (17)

is the open loop transfer function of a RTOB based force control system; $\varphi(s) = J_m \hat{K}_\tau s(s+\alpha g_{DOB}) + \hat{K}_\tau (D_{env}s+K_{env}) - \hat{J}_m K_\tau s(s+\beta g_{DOB})$; $\hat{J}_m = J_{mn} + \Delta\hat{J}_m$ and $\hat{K}_\tau = K_{\tau n} + \Delta\hat{K}_\tau$ are the identified inertia and torque coefficient, respectively; and $\beta = \dfrac{J_{mn}\hat{K}_\tau}{\hat{J}_m K_{\tau n}}$. The other parameters are same as defined above.

If a RTOB is designed by using perfect system identification, i.e., $\alpha = \beta$, then the open-loop transfer function is

$$L_{RTOB}(s) = C_f \frac{g_{RTOB}J_m \alpha(s+g_{DOB})(D_{env}s+K_{env})}{s\{J_m s(s+\alpha g_{DOB})+(D_{env}s+K_{env})\}(s+g_{RTOB})} \quad (18)$$

In general, the bandwidths of a DOB and a RTOB are set to the same value in a RTOB based robust force control system. If it is applied into (18), i.e., $g_{DOB} = g_{RTOB} = g$, then the open-loop transfer function is

$$L_{RTOB}(s) = C_f \frac{gJ_m \alpha(D_{env}s+K_{env})}{s\{J_m s(s+\alpha g)+(D_{env}s+K_{env})\}} \quad (19)$$

Equations (17), (18) and (19) show that each of the open loop transfer functions has a pole at the origin, so the steady state error of force control is removed by a DOB.

Let us start to analyze the robust force control system by considering (19). The relative degree of the open loop transfer function is two, so its root loci have asymptotes, at ±90°. The stability of the robust force control system deteriorates as the stiffness/damping of environment increases/decreases, since the zero of the open-loop transfer function at $-K_{env}/D_{env}$ moves away from the origin and the phase-lag increases.

If the bandwidths of DOB and RTOB are set to different values, then the relative degree of the open loop transfer function, i.e., the asymptotic behave of the root locus, does not change; however, a new phase lead-lag compensator, which can be used to improve the stability and performance, is obtained. It is shown clearly by rewriting (18) as follows:

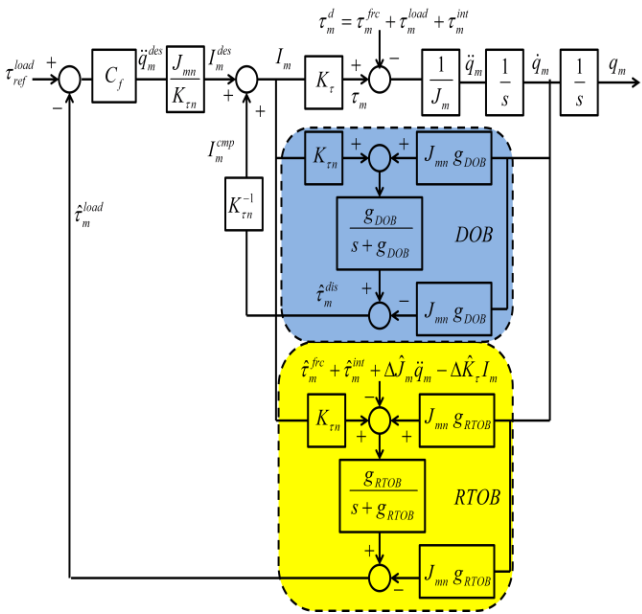

Fig. 4. A Block diagram of a RTOB based robust force control system

$$L_{RTOB}(s) = C_f C_{com} \frac{g_{RTOB} J_m \alpha (D_{env} s + K_{env})}{s\{J_m s(s + \alpha g_{DOB}) + (D_{env} s + K_{env})\}} \quad (20)$$

where $C_{com} = \frac{(s + g_{DOB})}{(s + g_{RTOB})}$ is a phase lead-lag compensator. The stability and performance of a RTOB based robust force control system can be improved by using $C_{com}$ as a phase lead compensator, i.e., $g_{RTOB} > g_{DOB}$.

So far, it is assumed that a RTOB is designed by using the perfect identification of inertia and torque coefficient. However, in practice, the identification of inertia is one of the most challenging issues in the design of a RTOB. Equation (17) shows that if an imperfect identification is used in the design of a RTOB, then the relative degree of the open-loop transfer function is one, so its root loci have asymptotes, at 180°. Although the asymptotic behave of the root locus improves by decreasing the relative degree of the open loop transfer function, the stability changes significantly by the imperfect identification. It can be explained as follows:

Let us consider the numerator of (17) by using

$$C_f g_{RTOB} \frac{J_{mn}}{K_{\tau n}} (s + g_{DOB}) \varphi(s) \quad (21)$$

where $\varphi(s) = (J_m \hat{K}_\tau - \hat{J}_m K_\tau) s^2 + \hat{K}_\tau D_{env} s + \hat{K}_\tau K_{env}$.

As it is shown in (21), $L_{RTOB}(s)$ has a right half plane (RHP) zero if $\hat{J}_m K_\tau > J_m \hat{K}_\tau$, i.e., $\alpha > \beta$. It is obvious that the stability and performance of the robust force control system deteriorate by the RHP zero. To overcome this issue, i.e., to obtain a minimum phase robust force control system, a new design constraint is proposed as follows:

$$\beta \geq \alpha \quad (22)$$

Consequently, the following design constraints should be considered in the design of DOB based robust motion control systems.

- The stability of a DOB based robust motion control system can be improved by increasing/decreasing the nominal inertia/torque coefficient.
- The peaks of the inner-loop's sensitivity and co-sensitivity transfer functions can be decreased, i.e., the robustness of a DOB can be improved, by decreasing/increasing the nominal inertia/torque coefficient. However, the stability deteriorates.
- In general, $\alpha = 2$ and $g_{DOB} \leq 0.25 g_v$ are useful design parameters to improve the stability and robustness.
- Setting $g_{RTOB} > g_{DOB}$ improves the stability of a RTOB based robust force control system.
- As $\alpha - \beta$ increases, the stability of a RTOB based robust force control system deteriorates and the bandwidth of the force control system gets lower. As a result, not only the performance, but also the stability deteriorates.
- As $\beta - \alpha$ increases, the stability of a RTOB based robust force control system improves; however, the performance deteriorates due to the estimation error of environmental impedance.

TABLE I
SIMULATION PARAMETERS

| Parameters | Descriptions | Values |
|---|---|---|
| Jmn | nominal motor inertia | 0.1 kgm$^2$ |
| $K_{\tau n}$ | nominal motor thrust coefficient | 5 Nm/A |
| $K_P$ | proportional gain of position control | 900 |
| $K_D$ | derivative gain of position control | 100 |

- To improve the performance of a RTOB based robust force control system, torque coefficient should be identified precisely in the design of a RTOB, yet the inertia identification can be neglected in many cases due to low accelerations in force control

IV. SIMULATION AND EXPERIMENT

In this section, simulation and experimental results will be presented.

*A. Simulation*

The parameters of the simulations are shown in Table I. Let us start by considering the robustness of a DOB. Fig.5 shows the co-sensitivity functions' frequency responses of the inner and outer loops, i.e., $T_{CoSEN}^{DOB}$ and $T_{CoSEN}^{PC}$, when a PD controller is implemented to achieve position control goals. As shown in Fig. 5a, the frequency responses of $T_{CoSEN}^{DOB}$ change significantly

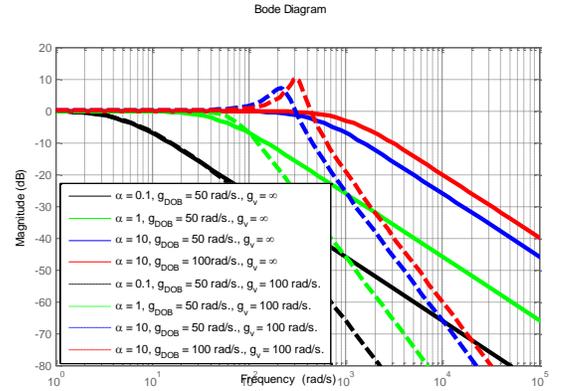

a) Inner-loop co-sensitivity function frequency responses

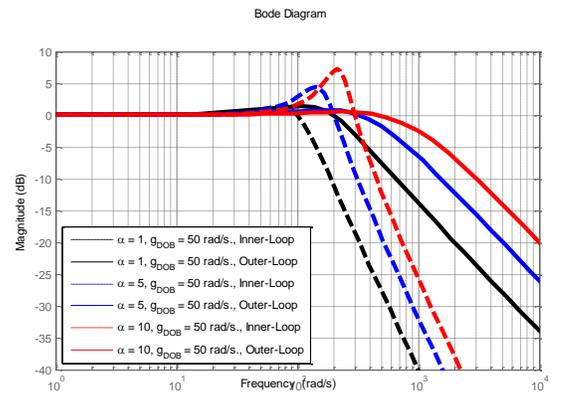

b) Outer-loop co-sensitivity function frequency responses

Fig. 5. Robustness of a DOB and a DOB based Position Control System

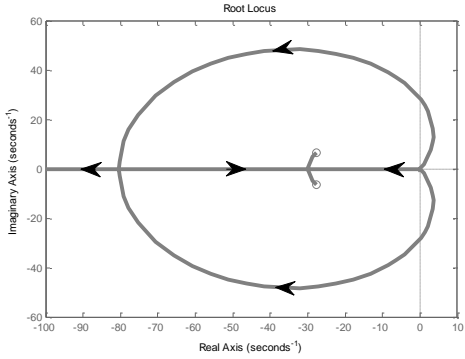

Fig. 6. Stability of a DOB based robust position control system

at high frequencies as $\alpha$ and/or $g_{DOB}$ are increased when $g_v$ is finite. Against the ideal velocity measurement case, $\alpha g_{DOB}$ cannot be increased freely due to the robustness constraint when practical velocity measurement is considered. However, the outer-loop position controller improves the robustness of the position control system as shown in Fig. 5b. Although the robustness of the outer-loop is improved by the performance controller, a DOB becomes more sensitive at high frequencies in the inner-loop as $\alpha g_{DOB}$ is increased.

Let us now consider the stability of a DOB based robust position control system. The root-locus of the robust position control system, which is shown in Fig. 6, is plotted with respect to $\alpha$ when $g_{DOB}$ is $500\,rad/s$. It shows that the stability of the robust position control system improves as $\alpha$ is increased. However, $\alpha$ is limited by the robustness constraint

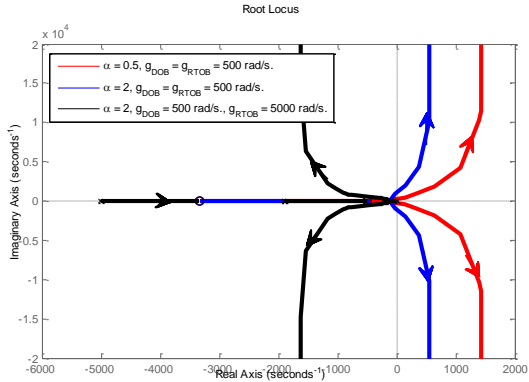

(a) $\alpha = \beta$

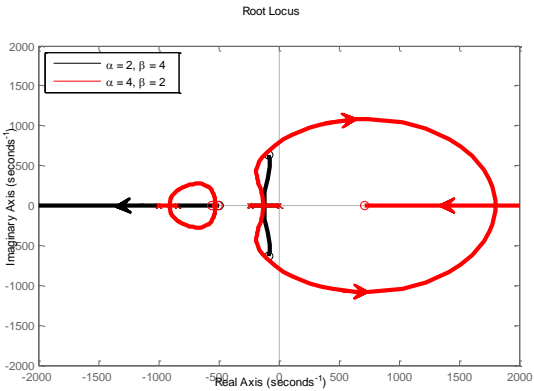

(b) $g_{DOB} = 500\,rad/s$ and $g_{RTOB} = 1000\,rad/s$

Fig. 7. Stability of a RTOB based robust force control system

TABLE II
EXPERIMENT PARAMETERS

| Parameters | Descriptions | Values |
| --- | --- | --- |
| $J_{m1}$ | inertia of motor 1 | 0.003 kgm$^2$ |
| $J_{m2}$ | inertia of motor 2 | 0.0003 kgm$^2$ |
| $K_\tau$ | motor thrust coefficient | 0.0603 Nm/A |
| $g_v$ | cut-off frequency of velocity measurement | 250 rad/s. |
| $K_P$ | proportional gain of position control | 400 |
| $K_D$ | derivative gain of position control | 45 |
| $C_f$ | proportional gain of force control | 1 |

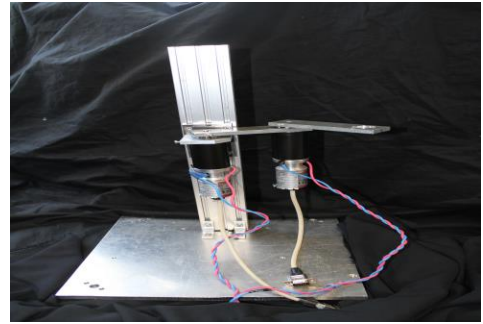

Fig. 8. Experimental setup: Two link planar arm

of a DOB, so there is a trade-off between the stability of the position control system and the robustness of a DOB.

Lastly, let us consider the stability of the robust force control system. The root-loci of a RTOB based robust force control system are plotted with respect to $C_f$ in Fig.7. Fig. 7a indicates that increasing $g_{RTOB}$ and/or $\alpha$ improves the stability of the robust force control system when the inertia and torque coefficient are identified precisely. However, $\alpha$ is limited by the robustness of a DOB, so there is a trade-off between the stability and robustness in the RTOB based robust force control systems as well. Fig. 7b indicates that the imperfect identification changes the stability of the robust force control system significantly: if $\alpha > \beta$, then the stability of the robust force control system deteriorates due to the RHP zero; however, if $\alpha < \beta$, then the stability is improved.

*B. Experiment*

A joint space control of a two-link planar robot arm, which is shown in Fig. 8, is carried out in the experiments. Specifications of the experimental setup are shown in Table II.

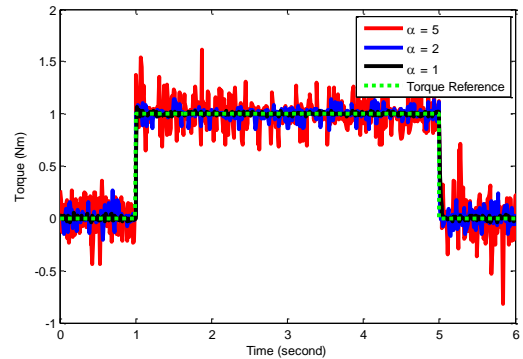

Fig. 9. Robustness / Noise suppression in the first motor

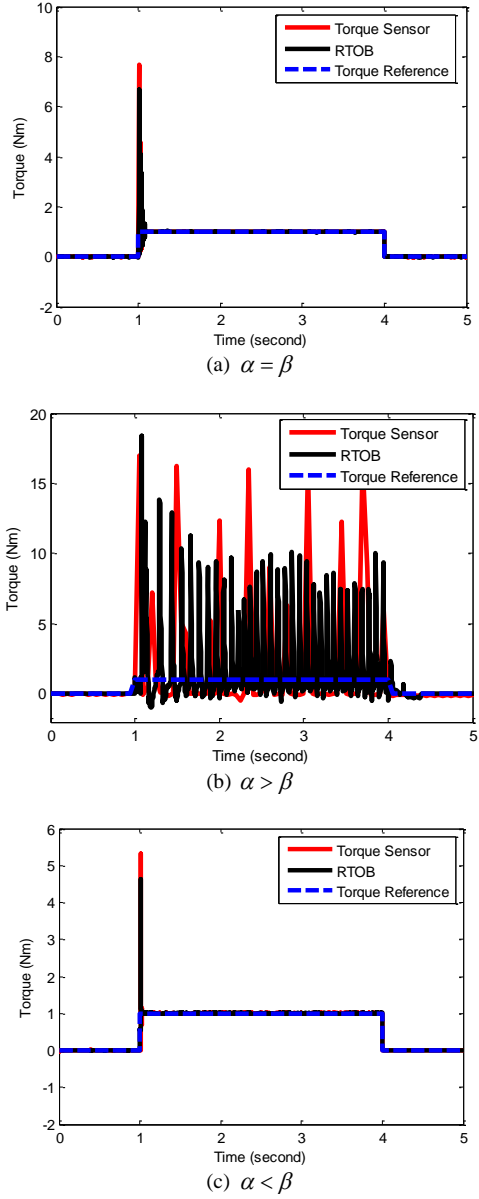

Fig. 10. Torque control responses of the first motor. $g_{DOB} = 75\,rad/s$ and
$$g_{RTOB} = 150\,rad/s$$
The sampling time is 0.1 ms.

Let us start by considering the robustness constraint of a DOB. Fig. 9 shows the torque control responses of the first motor when $\alpha$ has different values and $g_{DOB} = 200\,rad/s$. It is clear from the figure that as $\alpha$ is increased, DOB becomes more sensitive to noise since the robustness deteriorates. To improve the robustness of DOB, $\alpha$ and $g_{DOB}$ should be tuned by considering the robustness constraint given in (7).

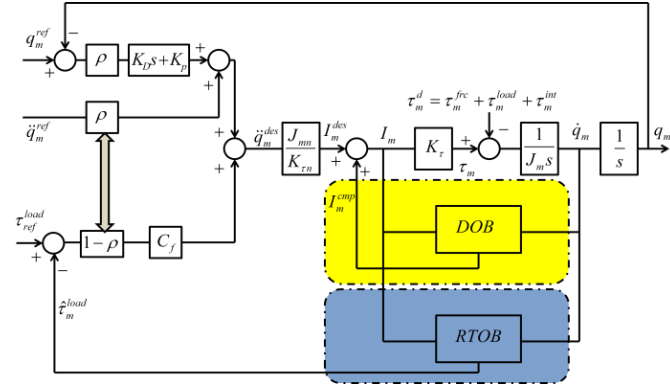

Fig. 11. Block diagram of acceleration control based hybrid motion control system

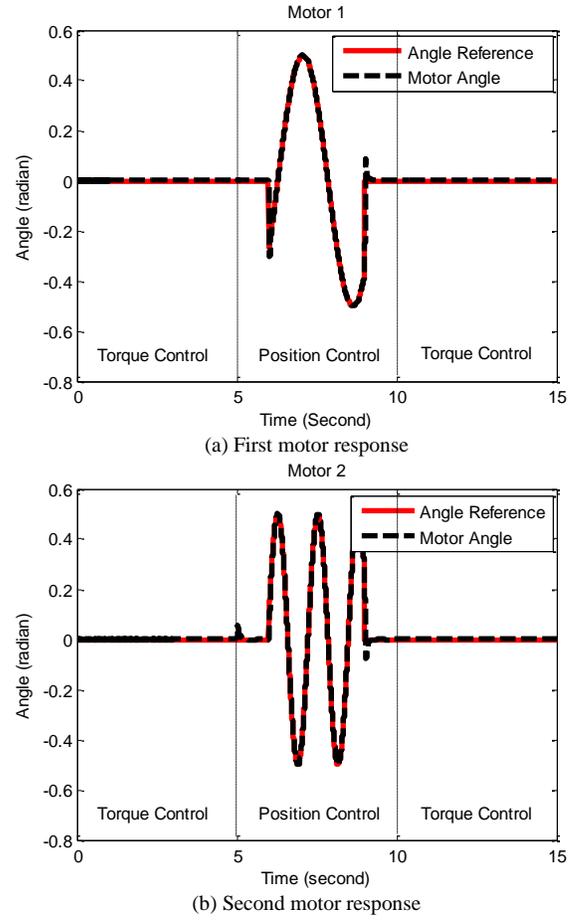

Fig. 12 Position control responses of the two link planar arm

Let us now consider the stability of the robust force control system. Fig. 10 shows the torque control responses of the first motor when the robot does not contact with environment initially. The second motor is controlled by using zero position control, and a step torque reference is applied to the first motor at 1 s. Fig. 10 clearly shows that the stability of the robust force control system changes significantly by the design parameters of DOB and RTOB and is improved by designing $\alpha \leq \beta$. To improve the performance of the robust force control system, $\alpha < \beta$ is designed by using $\hat{K}_\tau \cong K_\tau$ and $\hat{J}_m < J_m$.

Lastly, an acceleration control based hybrid motion control implementation is conducted by using the proposed position and force control systems. Block diagram of the acceleration control based hybrid motion control system is shown in Fig 11. In this figure $\rho$ denotes compliance selection constant. Torque control is conducted between 0 and 5 and 10 and 15 seconds; and position control is conducted between 5 and 10 seconds. Step torque reference inputs are applied to each joints at different times during the torque control, and the links interact with the environments initially; sinusoidal position reference inputs are applied to each joints, and plant uncertainties are identified to improve the performance and stability by using an online identification algorithm during the position control [26]. Fig. 12 and 13 show the position and torque control responses at each joints, respectively. As shown in the figures, the position and torque control goals can be achieved when the proposed methods are used. Besides, as shown in Fig. 13, the performance of the robust force control system i.e., RTOB, can be improved by identifying system uncertainties, such as friction, during position control.

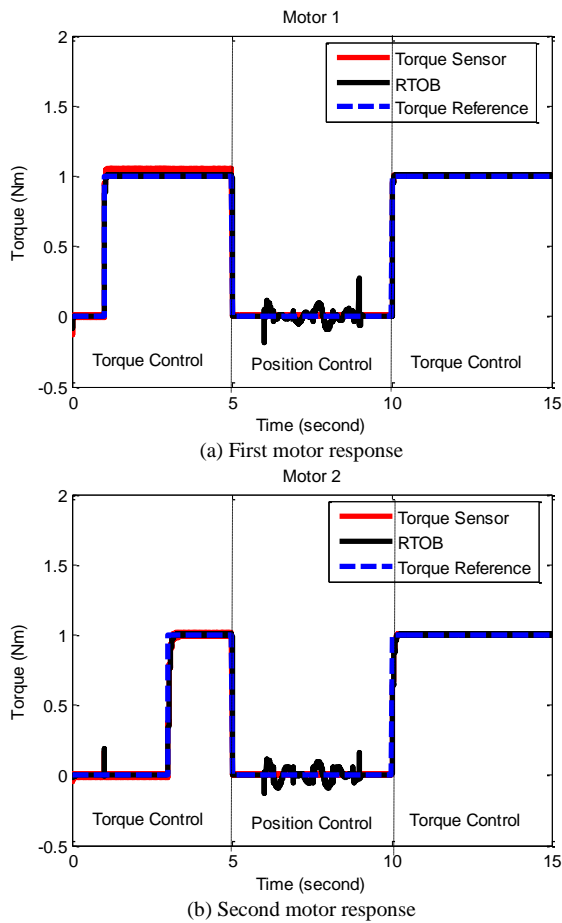

Fig. 13 Torque control responses of the two link planar arm

## V. CONCLUSION

This paper proposes new design tools for the DOB based robust motion control systems. It is shown that as the nominal inertia/torque coefficient is increased / decreased, the stability of the robust motion control system is improved, yet the robustness of a DOB deteriorates; and vice versa. A new design method which improves the stability and robustness of a DOB based motion control system is proposed. It is clear from the analyses that velocity measurement is of great importance for the stability, robustness and performance of a DOB based motion control system.

A new stability analysis is proposed for the RTOB based robust force control systems. It is shown that a DOB and a RTOB can be designed as a phase lead-lag compensator, and increasing the bandwidth of RTOB improves the stability of the robust force control system. When the inertia and torque coefficient cannot be identified precisely, the stability of the robust force control system changes significantly. If the identified inertia/torque coefficient is lower/higher than the exact one, then the stability of the robust force control system is improved; however, if the identified inertia/torque coefficient is higher/lower than the exact one, then the robust force control system has a right half plane zero, so its stability deteriorates significantly by increasing the force control gain. Therefore, not only the performance, but also the stability is affected by the identification errors in the design of a RTOB. To improve the performance of a RTOB based robust force control system, torque coefficient identification is crucial; however, inertia identification can be neglected in many cases due to low acceleration in force control. Therefore, if lower identified inertia is used in the design of a RTOB, then the stability of the robust force control system can be improved without degrading the performance. The simulation and experimental results show the validity of the proposals clearly.

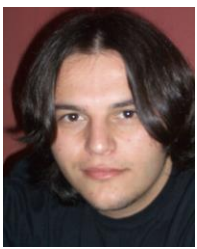
**Emre Sariyildiz** (S'11) received his Ms.C. degree in Mechatronics Engineering from Istanbul Technical University, Istanbul, Turkey in 2009. He is currently working toward the Ph.D. degree in integrated design engineering at Keio University, Yokohoma, Japan.

His main research interests are control theory and robotics.

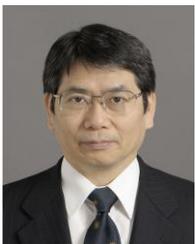
**Kouhei Ohnishi** (S'78–M'80–SM'00–F'01) received the B.E., M.E., and Ph.D. degrees in electrical engineering from the University of Tokyo, Tokyo, Japan, in 1975, 1977, and 1980, respectively.

Since 1980, he has been with Keio University, Yokohama, Japan. His research interests include mechatronics, motion control, robotics, and haptics.

Dr. Ohnishi received Best Paper Awards from the IEEJ and the Japan Society for Precision Engineering, and Outstanding Paper Awards at IECON'85, IECON'92, and IECON '93. He also received the EPE-PEMC Council Award and the Dr.-Ing. Eugene Mittelmann Achievement Award from the IEEE Industrial Electronics Society in 2004.